\documentclass[manuscript]{aastex}

\usepackage{natbib,aas_macros,amsmath}
\usepackage{times}
\usepackage{ulem}
\usepackage{multirow,color}

\begin{document}
\shorttitle{
Near-Infrared Unidentified Emission Bands in Galactic H\,{I\hspace{-.1em}I} regions
}
\shortauthors{Mori et al.}
%\slugcomment{Ver. December 15, 2012}
\slugcomment{in prep.}

\title{%
Observational studies on the Near-Infrared Unidentified Emission Bands in \\
Galactic H\,{I\hspace{-.1em}I} regions
}

\author{%
Tamami I. Mori\altaffilmark{1}, Takashi Onaka\altaffilmark{1}, Itsuki Sakon\altaffilmark{1}, 
\affil{Department of Astronomy, Graduate School of Science, The University of Tokyo}
Daisuke Ishihara\altaffilmark{2}, 
\affil{Graduate School of Science, Nagoya University}
Takashi Shimonishi\altaffilmark{3},
\affil{Department of Earth anf Planetary Sciences, Faculty of Science, Kobe University}
Ryou Ohsawa\altaffilmark{1}
\and
Aaron C. Bell\altaffilmark{1}
\affil{Department of Astronomy, Graduate School of Science, The University of Tokyo}
}

\email{morii@astron.s.u-tokyo.ac.jp}

\altaffiltext{1}{
Department of Astronomy, Graduate School of Science, The University of Tokyo, 
7-3-1 Hongo, Bunkyo-ku, Tokyo 113-0033, Japan
}
\altaffiltext{2}{
Graduate School of Science, Nagoya University, 
Chikusa-ku, Nagoya 464-8602, Japan
}
\altaffiltext{3}{
Department of Earth anf Planetary Sciences, Faculty of Science, Kobe University, 
1-1 Rokkodai-cho, Nada Kobe 657-8501 Japan}

%---------------------------------------------------------------------
\begin{abstract}
Using a large collection of near-infrared spectra (2.5--5.4\,$\mu$m) 
of Galactic H\,{I\hspace{-.1em}I} regions and H\,{I\hspace{-.1em}I} region-like objects,  
we perform a systematic investigation of the astronomical polycyclic aromatic hydrocarbon (PAH) features. 
36 objects were observed 
by the use of the infrared camera onboard the {\it AKARI} satellite 
as a part of a directer's time program. 
In addition to the well-known 3.3--3.6\,$\mu$m features, 
most spectra show a relatively-weak emission feature at 5.22\,$\mu$m with sufficient signal-to-noise ratios, 
which we identify as the PAH 5.25\,$\mu$m band previously reported. 
By careful analysis, we find good correlations between the 5.25\,$\mu$m band and 
both the aromatic hydrocarbon feature at 3.3\,$\mu$m and the aliphatic ones at around 3.4--3.6\,$\mu$m. 
The present results give us convincing evidence 
that the astronomical 5.25\,$\mu$m band is associated with C-H vibrations as suggested by previous studies 
and show its potential to probe the PAH size distribution. 
The analysis also shows that the aliphatic to aromatic ratio of $I_{\rm 3.4-3.6\,\mu m}/I_{\rm 3.3\,\mu m}$ decreases against the ratio of 
the 3.7\,$\mu$m continuum intensity to the 3.3\,$\mu$m band, $I_{\rm cont, 3.7\,\mu m}/I_{\rm 3.3\,\mu m}$, 
which is an indicator of the ionization fraction of PAHs. 
The mid-infrared color of $I_{\rm 9\,\mu m}/I_{\rm 18\,\mu m}$ also declines steeply against the ratio of 
the hydrogen recombination line Br$\alpha$ at 4.05\,$\mu$m to the 3.3\,$\mu$m band, 
$I_{\rm Br\alpha}/I_{\rm 3.3\,\mu m}$. 
These facts indicate possible dust processing inside or at the boundary of ionized gas. 
\end{abstract}

%---------------------------------------------------------------------
%---------------------------------------------------------------------
%---------------------------------------------------------------------

%%%%%%%%%%%%%%%%%%%%%%%%%%%%%%%%%%
%%%%%%%%%%%%%%%%%%%%%%%%%%%%%%%%%%
\section{Introduction \label{sec:introduction}}
%%%%%%%%%%%%%%%%%%%%%%%%%%%%%%%%%%
%%%%%%%%%%%%%%%%%%%%%%%%%%%%%%%%%%

H\,{I\hspace{-.1em}I} regions are luminous gas clouds ionized by embedded massive O-type or B-type stars. 
Considering their extreme physical conditions (intense radiation fields and stellar winds from the central stars), 
H\,{I\hspace{-.1em}I} regions are a good laboratory for investigating the evolution of materials in the interstellar medium (ISM). 
It can be assumed that destruction and/or processing of dust grains occurs in the interior or boundary of H\,{I\hspace{-.1em}I} regions, 
but the details remain to be explored. 

The near-infrared (NIR) wavelength range of 2--5\,$\mu$m includes a number of emission and/or absorption features 
related to various kinds of gaseous and solid materials in the ISM, 
containing significant pieces of information about the interstellar environment and the evolution of cosmic material. 
Over the past few decades, 
several space satellite missions such as the {\it IRAS}, {\it IRTS}, {\it ISO}, and {\it Spitzer} 
have explored the universe at infrared (IR) wavelengths and demonstrated its wealth. 
However, due to observational constraints, 
the spectral range shorter than 5.5\,$\mu$m has not come under the spotlight.  
The {\it ISO} SWS spectrometer provides spectroscopic information on this range, 
but they are limited only for very bright sources. 
Since the advent of the AKARI telescope \citep{mura07}, 
which has higher sensitivity than ever before at NIR wavelengths, 
the richness of this wavelength range has been gradually recognized. 
Among the spectral features appearing in this NIR range, 
a series of IR emission features associated with the smallest variety of carbonaceous species in the ISM, 
which are called ``the Unidentified infrared (UIR) bands'', 
are one of the most interesting and ubiquitous features in space \citep[e.g.][]{gill73,onak96,peet02,lutz05}. 
Although the exact band carriers are not identified yet, 
they are generally ascribed to 
the infrared fluorescence of polycyclic aromatic hydrocarbons \citep[PAHs, e.g.][]{tiel08,jobl11,lidr12} 
or related hydrocarbon materials \citep[e.g.][]{saka84,pule89,papo89,alla89-a,kwok11}. 
Since their spectral properties vary with the host environment,
these IR emission features have been enthusiastically studied from various perspectives 
as a possible probe of the ISM conditions and star formation history. 
In particular, the 3.3--3.6\,$\mu$m region is diagnostically important 
in identifying the aromatic-aliphatic nature of the band carriers (hereafter symbolically called PAHs). 
The strong emission band peaking at 3.3\,$\mu$m is assigned to a stretching mode of aromatic C-H bonds, 
whereas the adjacent 3.4--3.6\,$\mu$m sub-features are thought to be due to vibrational modes of aliphatic C-H bonds \citep[]{dule81}, 
or to arise from overtone ($n\times\nu_{i}$) bands of the 3.3\,$\mu$m fundamental aromatic C-H stretching mode. 
It is also known that astronomical spectra exhibit a weak emission feature around 5.25\,$\mu$m \citep{alla89-b}. 
Previous studies suggest that this feature is attributed to a mixture of overtone, difference ($\nu_{i}-\nu_{j}$), and combination ($\nu_{i}+\nu_{j}$) bands of fundamental frequencies of stretching and bending vibrations of aromatic C-H bonds \citep{alla89-b,boer09}. 
However, because of its weakness, 
there is only limited observational data and a detailed observational analysis has not been carried out for the 5.25\,$\mu$m band. 

This paper presents the observational results of NIR slit spectroscopy 
of 36 Galactic H\,{I\hspace{-.1em}I} region or H\,{I\hspace{-.1em}I} region-like objects 
with the infrared camera (IRC) onboard {\it AKARI} \citep{onak07}. 
The 5.25\,$\mu$m band appears at the edge of the spectral range of the prism spectroscopy with {\it AKARI}/IRC, 
which enables us to study this minor PAH band for much more samples than previous studies. 
We first report the detection of this minor PAH band at 5.25\,$\mu$m with sufficient S/N ratios. 
We also discuss the nature and evolution of dust grains in the H\,{I\hspace{-.1em}I} region environment 
based on the variation of 
the aromatic to aliphatic ratio, $I_{\rm 3.4-3.6\,\mu m}/I_{\rm 3.3\,\mu m}$, and 
the MIR color, $I_{\rm 9\,\mu m}/I_{\rm 18\,\mu m}$.

Section \ref{sec:observations} describes the observations and data reduction. 
Spectral analysis is presented in Section \ref{sec:analysis}, 
and its results and discussion in Section \ref{sec:discussion}. 
Summary and conclusion are given in Section \ref{sec:summary}.

%%%%%%%%%%%%%%%%%%%%%%%%%%%%%%%%%%
%%%%%%%%%%%%%%%%%%%%%%%%%%%%%%%%%%
\section{Observations} \label{sec:observations}
%%%%%%%%%%%%%%%%%%%%%%%%%%%%%%%%%%
%%%%%%%%%%%%%%%%%%%%%%%%%%%%%%%%%%

%%%%%%%%%%%%%%%%%%%%%%%%%%%%%%%%%%
\subsection{NIR slit spectroscopy with the AKARI/IRC} \label{sec:observations:1}
\begin{figure*}
\begin{center}
\includegraphics[scale=1.75]{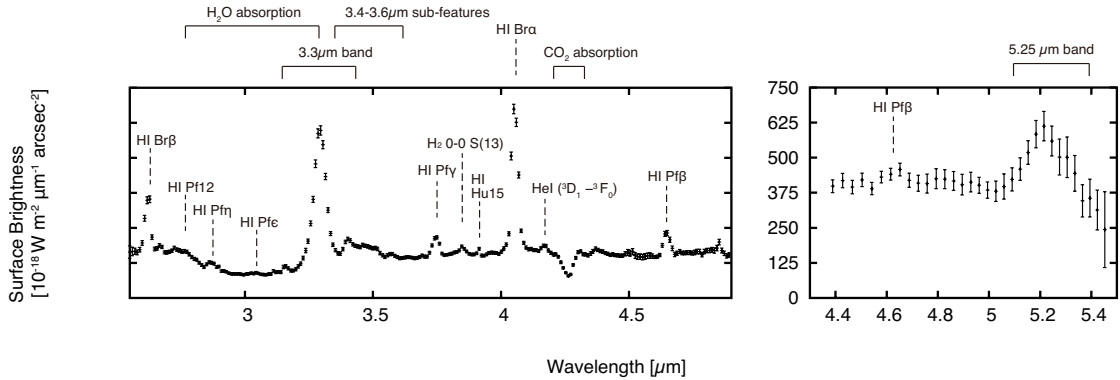}
\caption{
A typical example of the spectra obtained by the present observations. 
They are taken from the position ``-04" of W31a (ID: 5200165.1) by the use of the ``Nh''
 slit. 
The left panel indicates the spectra taken with the grism ($R$\,$\sim$\,100), 
and the right panel that of the prism ($R$\,$\sim$\,20--40). 
}
\label{fig:1}
\end{center}
\end{figure*}

%%%%%%%%%%%%%%%%%%%%%%%%%%%%%%%%%%
The present studies are mainly  based on 
the slit spectroscopic data obtained by the {\it AKARI}/IRC 
in the framework of director's time program 
during the {\it AKARI} post-helium mission phase (Phase 3), 
during which only NIR channel (2.0--5.5\,$\mu$m) was in operation. 
All of the observations were performed by the use of the ``Ns''
 or ``Nh''
 slit 
(0.8$^{\prime}$\,length\,$\times$\,5$^{\prime\prime}$ width or 1$^{\prime}$\,length\,$\times$\,3$^{\prime\prime}$\,width, respectively) 
for diffuse sources (see \cite{onak07} for more detailed information on the instrument). 
The director's time program was planned for the purpose of calibration. 
In the present observations, 
for the the relative calibration of the NIR disperser, 
grism and prism spectroscopy were done separately 
for the same slit area 
in the fist and the latter half of a single pointing observation. 
This procedure provides two kinds of spectra for the same slit area 
with different wavelength coverages and spectral resolutions 
(2.5--5.0\,$\mu$m, $R$\,$\sim$\,100 for the grism and
 1.7--5.4\,$\mu$m, $R$\,$\sim$\,20--40 for the prism, 
see Figure \ref{fig:1}).  
The target sources were selected from 
Galactic ultra-compact (UCH\,{I\hspace{-.1em}I}) and giant H\,{I\hspace{-.1em}I} (GH\,{I\hspace{-.1em}I}) regions catalogued by \cite{crow03} and \cite{cont04} 
according to the visibility of the satellite, 
except for two unclassified  infrared sources, G331.386-0.359 and G345.528-0.051. 
These sources are unclassified,
but their spectra show H\,{I\hspace{-.1em}I} region-like properties, and 
we classify them as H\,{I\hspace{-.1em}I} region-like objects. 
Generally, two pointing observations were performed for each target, 
whereas some targets were observed only once. 
We selected 61 pointing observations of 36 different targets in total 
that are not severely affected by instrumental artifacts or stray light. 
The observation logs and target parameters are summarized in Table \ref{tab:1}. 
We narrowed the selection further at the spectral analysis stage, 
by removing those without a sufficient signal-to-noise ratio to quantify the PAH emission features and extinction, 
and in some cases those without available complementary MIR imaging data (see details below) 
from the following analysis. 
The accurate slit position is determined from the 3.2\,$\mu$m image (N3) taken during each pointing observation, 
by referring to the positions of point sources in the 2MASS catalog.

%%%%%%%%%%%%%%%%%%%%%%%%%%%%%%%%%%
\subsection{Data reduction} \label{sec:observations:2}
%%%%%%%%%%%%%%%%%%%%%%%%%%%%%%%%%%
The data reduction basically follows the official IDL pipeline for {\it AKARI} Phase 3 data \citep[ver.20111121,][]{ohya07,onak09b}. 
Since the wavelength is fixed for a given detector column in the IRC slit spectroscopy, 
a systematic pattern from the detector anomaly or a dark pattern in the column direction sometimes persists 
even after the standard pipeline process. 
The grism spectroscopic data are especially sensitive to the fluctuation of those patterns. 
Therefore, we perform additional dark-subtraction for the grism spectroscopic data. 
In the grism mode, the brink of the image can be used to correct for some systematic error. 
It is blocked by the aperture mask and allows us to estimate the residual dark currents. 
We obtain median values of 10 pixels corresponding to a given wavelength in this region, and subtract them from the spectrum. 

The spectral data generally show a spatial variation within the entire slit area. 
The FWHM of the point-spread function of the IRC 
is approximately 3.2 pixels ($\sim$4.6$^{\prime\prime}$) at the NIR wavelengths during Phase 3 period \citep[]{onak10}. 
We extract spectra taken with the grism and the prism mode respectively 
from the same slit area with the length of 6 pixels ($\sim$9$^{\prime\prime}$) one by one along the slit direction, 
avoiding those affected by point-like sources. 
As a result, we obtain several sets of grism and prism spectra from each pointing observation. 
% (see Appendix B)
Even though the telescope is pointed towards the same position on the sky, 
due to the different optical alignments, 
spectral images are shifted systematically by about 4 pixels along the slit direction between the grism and the prism modes. 
When we create a pair of grism and prism spectra of the same position, 
this position shift is taken into consideration. 
Figure \ref{fig:1} gives an example of the obtained spectra. 
In the short wavelength region, 
the spectral resolution of the prism is too low to discuss features quantitatively. 
Furthermore, saturation frequently occurs in the prism spectra 
because of its extremely high sensitivity 
particularly in this wavelength range. 
For these reasons, we truncate the prism spectra, 
and only use a 4.35--5.4\,$\mu$m wavelength region in the following analysis. 
All of the resultant spectra are given in the published version 
together with the individual {\it S9W} band images taken by the {\it AKARI} MIR all-sky survey (see details below).
%, in which the location of the slit and each spectra-extracted region are depicted with colored boxes. 

%%%%%%%%%%%%%%%%%%%%%%%%%%%%%%%%%%
\subsection{Ancillary MIR Imaging data} \label{sec:observations:3}
%%%%%%%%%%%%%%%%%%%%%%%%%%%%%%%%%%
We secondarily utilize the diffuse data obtained in the {\it AKARI} MIR all-sky survey \citep{ishi10}. 
The {\it AKARI} MIR all-sky survey was carried out by the IRC at two MIR wide band filters, 
{\it S9W} ($\lambda_{\rm eff}$=9\,$\mu$m, 6.7--11.6\,$\mu$m) and 
{\it L18W} ($\lambda_{\rm eff}$=18\,$\mu$m, 13.9--25.6\,$\mu$m). 
In the NIR regime, the effect of the zodiacal light is negligible, 
but its intensity increases dramatically by a factor of approximately one hundred in units of Jy at MIR wavelengths. 
We therefore subtract the medium value of a relatively dark area within the same frame from the processed images 
as the background, most of which is expected to arise from the zodiacal emission, 
and perform aperture-photometry within a 5$^{\prime\prime}$ radius circle centered on each spectrum-extracted region. 
From the nature of the present targets, 
the background signals that we subtract are considerably small relative to the source ones in most cases. 
The pixels where saturation occurs due to the presence of bright astronomical sources are automatically masked through the data reduction process. 
Some of the present target objects (e.g. M17a) contain such masked regions, 
which are excluded from the analysis involving the MIR imaging data.

%%%%%%%%%%%%%%%%%%%%%%%%%%%%%%%%%%
%%%%%%%%%%%%%%%%%%%%%%%%%%%%%%%%%%
\section{Spectral Analysis} \label{sec:analysis}
%%%%%%%%%%%%%%%%%%%%%%%%%%%%%%%%%%
%%%%%%%%%%%%%%%%%%%%%%%%%%%%%%%%%% 
As shown in Figure \ref{fig:1}, 
the obtained spectra are rich in ISM features, 
reflecting the complex geometry of the sources. 
A number of hydrogen and helium recombination lines 
(e.g. Br$\beta$ at 2.63\,$\mu$m, and Br$\alpha$ at 4.05\,$\mu$m) signify ionized gas, and 
the grism spectra clearly show the distinct 3.3\,$\mu$m PAH emission band 
and the adjacent weak sub-features around 3.4--3.6\,$\mu$m. 
Intriguingly, a broad emission feature can be seen 
in the unique coverage of the prism spectra of 5.0--5.4\,$\mu$m. 
The central wavelength is estimated as 5.22\,$\mu$m in the present spectra. 
It has been suggested that interstellar PAHs give a faint emission feature around this area \citep[e.g.][]{alla89-b,boer09}. 
\cite{boer09} report that the central wavelength of this PAH emission feature is 5.25\,$\mu$m. 
This value differs from that is measured from the present dataset by about 1 pixel ($\sim$0.03\,$\mu$m), 
but the discrepancy is still within the uncertainty of the wavelength calibration particularly because of the edge of the spectrum. 
Hydrogen recombination line Hu$\delta$ at 5.13\,$\mu$m is also a possible source of this feature. 
However, we can recognize that a line around at 5.13\,$\mu$m is perched on the tail of the feature 
in some spectra like those of RCW42 and RCW49, and Hu$\delta$ is clearly separated from the 5.22\,$\mu$m feature.  
In addition to Hu$\delta$, several hydrogen and helium recombination lines fall in the range of 5.0--5.4\,$\mu$m, 
but they are negligibly small when compared to Hu$\delta$. 
Therefore, we identify the observed feature as the PAH 5.25\,$\mu$m band, 
although we cannot completely rule out the possibility of some contamination. 
The uncertainty in the absolute flux calibration accompanied by the wavelength uncertainty 
is less than 5\%, and is within the observational error. 
In some cases, 
clear absorption features associated with H$_2$O and CO$_2$ ices 
also appear at around 3.05 and 4.27\,$\mu$m. 
These ice absorption features will be investigated in a separated paper (Mori et al. in prep), 
and are not discussed further here. 
We also find that 
in the present IRC spectra 
there is no apparent signature of deuterated PAHs (PADs) around 4.3--4.7\,$\mu$m \citep[e.g.][]{alla89-a,peet04-b}. 
The aromatic and aliphatic C-D stretching bands are known to appear in this wavelength range. 
The detailed analysis and discussion on these PAD features are presented in \cite{onak14}. 

\begin{figure*}
\begin{center}
\includegraphics[scale=1.75]{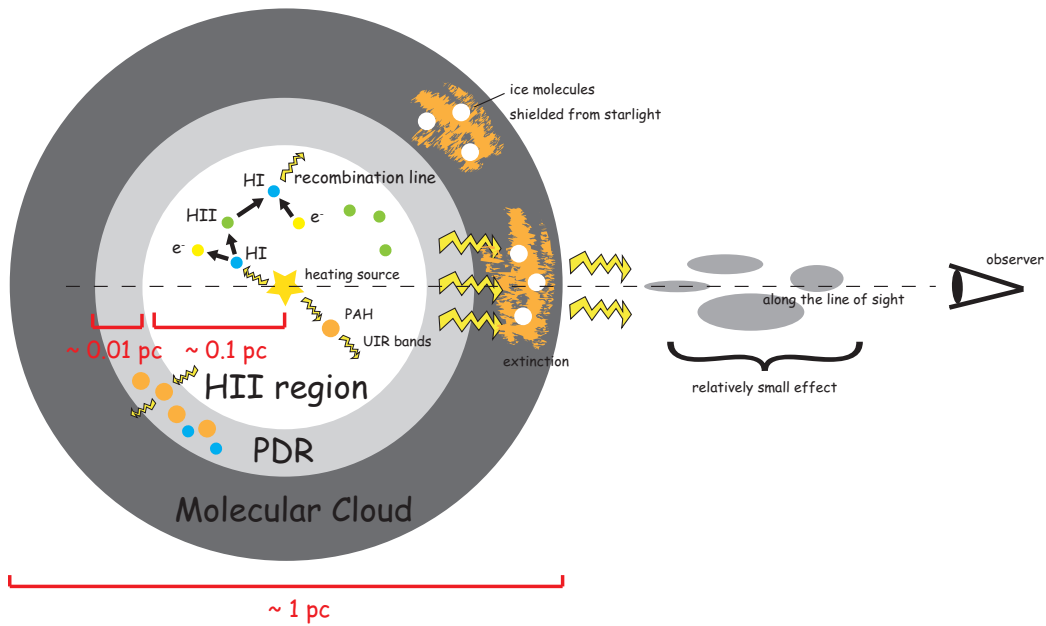}
\caption{
Illustration of the assumed configuration of a H\,{I\hspace{-.1em}I} region and its surroundings along the line of sight. 
 }
\label{fig:2}
\end{center}
\end{figure*}

Taking these distinctive spectral features into account, 
we use a least--squares method for the spectral fitting on the following supposition 
to investigate the observed features quantitatively. 
We assume that the emission features such as the hydrogen and helium recombination lines and the PAH emission features come entirely from H\,{I\hspace{-.1em}I} region--PDR complexes excited by embedded stars, and that those emissions are attenuated mostly by adjacent clouds, which contain and shield the ices from starlight. 
The schematic diagram of this assumed configuration is presented in Figure \ref{fig:2}.

%%%%%%%%%%%%%%%%%%%%%%%%%%%%%%%%%%
\subsection{Fitting for the grism spectra} \label{sec:analysis:1}
\begin{figure*}
\begin{center}
\includegraphics[scale=1.75]{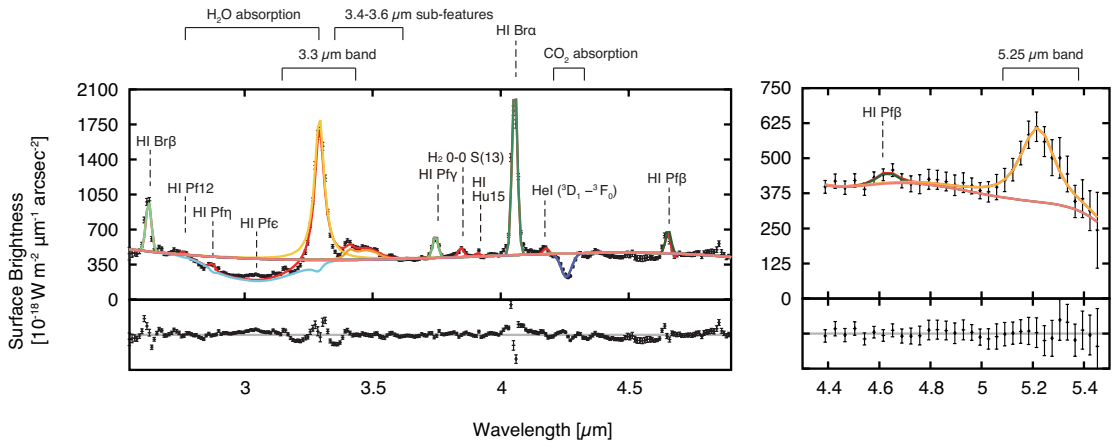}
\caption{
Fitting results for the spectra shown in Figure \ref{fig:1}. 
The salmon lines designate the continuum component of a quintic function. 
The major emission line features 
like Br$\beta$ at 2.63\,$\mu$m, Pf$\gamma$ 3.75\,$\mu$m, Br$\alpha$ 4.05\,$\mu$m, and Pf$\beta$ 4.65\,$\mu$m (green lines), 
the PAH emission bands (yellow lines), 
and the ice absorption features (blue lines) are respectively overlaid on the best-fit model spectra (red lines). 
As for the the grism spectrum, the wavelength region from 2.9 to 3.1\,$\mu$m is out of the spectral fitting (see details in \S \ref{sec:analysis:1}). 
The lower panels indicate the residual spectra of each plot. 
}
\label{fig:3}
\end{center}
\end{figure*}

%%%%%%%%%%%%%%%%%%%%%%%%%%%%%%%%%%
First, we discuss the analysis of spectra taken with the grism. 
The ice absorption features can be regarded as a ``foreground-screen''. 
Except for the broad absorption feature around 3.05\,$\mu$m associated with H$_2$O ice, the CO and CO$_2$ ice absorption features cannot be completely resolved with the spectral resolution of the {\it AKARI}/IRC slit spectroscopy. 
We fit the laboratory spectrum of the pure H$_2$O ice at 10\,K to the observed spectra, while the CO and CO$_2$ ice absorption features are modeled with Gaussian functions with negative signs. 
Then, the function we fit to the observed spectra is given by 
\begin{eqnarray}
\nonumber F_{{\scriptscriptstyle \lambda}}{\scriptstyle (\lambda)}\,{\scriptscriptstyle=}\biggl[  
   \sum_{\scriptscriptstyle k=0}^{\scriptscriptstyle 5}     a_{\scriptscriptstyle k}      \lambda^{\scriptscriptstyle k} 
{\scriptscriptstyle+} \sum_{\scriptscriptstyle k_l=1}^{\scriptscriptstyle 2}  b_{\scriptscriptstyle k_l}   f_{\scriptscriptstyle {\rm l}}{\scriptstyle (\lambda;\lambda_{\scriptscriptstyle k_l};\gamma_{\scriptscriptstyle k_l})}
\,{\scriptscriptstyle+} \sum_{\scriptscriptstyle k_g=1}^{\scriptscriptstyle 11}c_{\scriptscriptstyle k_g} f_{\scriptscriptstyle {\rm g}}{\scriptstyle (\lambda;\lambda_{\scriptscriptstyle k_g};\gamma_{\scriptscriptstyle k_g})} \biggr] \\
\cdot \exp(-d\,{\rm ln10}\,A_{\scriptscriptstyle \rm H_2O}{\scriptstyle ({\rm \lambda})})
\,{\scriptscriptstyle-} \sum_{\scriptscriptstyle k'_g=1}^{\scriptscriptstyle 2}e_{\scriptscriptstyle k'_g} f_{\scriptscriptstyle {\rm g}}{\scriptstyle (\lambda;\lambda_{\scriptscriptstyle k'_g};\gamma_{\scriptscriptstyle k'_g})},\,\,\,\,\,\,\,\,\,\,\,\,
\label{eq:grism_fit}
\end{eqnarray} 
where $f_{\rm l}$($\lambda;\lambda_{\rm 0};\gamma$) and $f_{\rm g}$($\lambda;\lambda_{\rm 0};\gamma$) denote Gaussian and Lorentzian functions with the central wavelength $\lambda_{\rm 0}$ and the FWHM $\gamma$ respectively. 
$A_{\rm H_2O}({\rm \lambda})$ denotes the absorbance of the H$_2$O ice. 
Here, we employ the laboratory data, 
which are retrieved from the Leiden atomic and molecular database \cite[]{ehre96}. 
The second term represents the CO and CO$_2$ absorption features. 
The FWHM $\gamma_{k'_g}$ and the center of the wavelength $\lambda_{k'_g}$ are fixed to the best-fit values, which are estimated from the spectra with good signal-to-noise ratios. 
Due to the different slit width (5$^{\prime\prime}$ for the ``Ns''
 slit and 3$^{\prime\prime}$ for the ``Nh''
 slit), the spectral resolution is slightly different between the spectra taken with ``Ns''
 slit and those with ``Nh''
 slit, and thus the best-fit values are derived for the ``Ns''
 slit and ``Nh''
 slit respectively (See Tables \ref{tab:2}). 
The terms in the first brackets of Eq. (1) represent emission features, which are made up of three components, a continuum, PAH, and line emission features. 
The continuum is modeled with a polynomial function of the 5th order and constrained to be non-negative. 
We express both the PAH 3.3\,$\mu$m and the adjacent 3.4\,$\mu$m sub-feature with a Lorentzian function, whereas the sub-features at around the 3.5\,$\mu$m are altogether modeled by one Gaussian function centered at 3.48\,$\mu$m, even though they are supposed to consist of multiple components. 
It is because the spectral resolution achieved by {\it AKARI}/IRC is not sufficient to resolve each of faint features around the 3.5\,$\mu$m. 
We attempt other combinations of Gaussians and Lorentzians, but the adopted combination provides the best fit. 
The center of the wavelength and the FWHM of these three components are fixed to the best-fit values estimated for the spectra obtained with {\it AKARI}/IRC (See Tables \ref{tab:3}).  
The emission lines are modeled with Gaussian profiles with the FWHM $\gamma_{k_g}$ fixed to match with the spectral resolution, 0.031\,$\mu$m for ``Ns''
 and 0.025\,$\mu$m for ``Nh''
 (See Table \ref{tab:4}). 
In the fitting, only $a_{k}$, $b_{k_l}$, $c_{k_g}$, $d$, and $e_{k'_g}$ are free parameters. 
The fitting is non-linear and thus we use the Levenberg--Marquardt method \citep[]{pres02}. 
We note that the wavelength region of 2.9--3.1\,$\mu$m is not used in the fitting, since this region is at the bottom of the H$_2$O absorption feature and is easily contaminated by weak foreground emissions. 
The left panel of Figure \ref{fig:3} shows an example of the fitting for the grism spectra.

%%%%%%%%%%%%%%%%%%%%%%%%%%%%%%%%%%
\subsection{Fitting for the prism spectra} \label{sec:analysis:2}
%%%%%%%%%%%%%%%%%%%%%%%%%%%%%%%%%%
In the prism spectroscopy, the spectral dispersion varies with wavelength. 
Taking into the account of this non-uniform spectral dispersion in the prism mode, 
we succeed in reproducing the prism spectrum from the grism spectrum by a simulation (see details in Appendix A). 
On the basis of this analytical result, 
we simulate the spectral profile in a spectrum taken with the prism mode, $f_{k_f}$($\lambda$), for each fitting component. 
As the fitting components, 
three emission features are selected; Pf$\beta$ at 4.65\,$\mu$m, Hu$\delta$ at 5.13\,$\mu$m, and the 5.25\,$\mu$m band, 
which are clearly seen in most of the resultant spectra taken with the prism. 
Then, we fit the spectra with the function 
\begin{eqnarray}
 F_{\scriptscriptstyle \lambda}{\scriptstyle (\lambda)}\,{\scriptscriptstyle=}
 \sum_{\scriptscriptstyle k=0}^{\scriptscriptstyle 5}a_{\scriptscriptstyle k} \lambda^{\scriptscriptstyle k} + \sum_{\scriptscriptstyle k_f=1}^{\scriptscriptstyle 3}b_{\scriptscriptstyle k_f} \cdot f_{\scriptscriptstyle k_f}{\scriptstyle (\lambda)},  
\label{fit}
\end{eqnarray}
where the first term represents the continuum, and the $f_{k_f}$($\lambda$) in the second term is the modeled function of each fitting component. 
In the formulation of the $f_{k_f}$($\lambda$), 
we assume that Pf$\beta$ and Hu$\delta$ lines have a Gaussian profile 
and the 5.25\,$\mu$m band has a Lorentzian profile in grism spectra. 
The FWHM of the 5.25\,$\mu$m band is set to the literature value reported by \cite{boer09}, 0.12\,$\mu$m, 
and the central wavelength is slightly modified to match with the best-fit value for the {\it AKARI}/IRC prism spectroscopy, 5.22\,$\mu$m. 
An example of the fits is given by the right panel of Figure \ref{fig:3}. 
As described above, the wavelength calibration in the prism mode is no better than $\sim$0.03\,$\mu$m in this spectral range, but the assumed spectral profile fits the observed spectrum fairly well.

%%%%%%%%%%%%%%%%%%%%%%%%%%%%%%%%%%
\subsection{Estimate of the extinction} \label{sec:analysis:3}
%%%%%%%%%%%%%%%%%%%%%%%%%%%%%%%%%%
We adopt the dust model of ``Milky Way, $R_V$=3.1'' of \cite{wedr01} as the extinction curve, and estimate the value of a visual extinction $A_V$ from the observed intensity ratio of Br$\beta$ to Br$\alpha$ in the assumption of the Case B condition of $T_{\rm e}=10^4\,{\rm K}$ and $n_{\rm e}=10^4\,{\rm cm^{-3}}$ \citep[]{sthu95} and the foreground-screen extinction (see Figure \ref{fig:2}). 
It is remarked that the line and band intensities derived from the fitting are corrected for the extinction. 
The $A_V$ value ranges widely from 0 to 40\,mag. 
In several regions, the effect of the extinction cannot be ignored. 
In addition to the extinction correction, 
we also correct for a contribution from the unresolved emission line Pf$\delta$ at 3.30\,$\mu$m to the UIR 3.3\,$\mu$m band. 
According to the Case B condition, we assume that the intensity of Pf$\delta$ is equal to 9.3\% of the extinction-corrected intensity of Br$\alpha$, and subtract it from the extinction-corrected intensity of the 3.3\,$\mu$m band.

%%%%%%%%%%%%%%%%%%%%%%%%%%%%%%%%%%
%%%%%%%%%%%%%%%%%%%%%%%%%%%%%%%%%%
\section{Discussion} \label{sec:discussion}
%%%%%%%%%%%%%%%%%%%%%%%%%%%%%%%%%%
%%%%%%%%%%%%%%%%%%%%%%%%%%%%%%%%%%
In most resultant spectra, as well as the 3.3\,$\mu$m band, 
both the 3.4--3.6\,$\mu$m sub-features and the 5.25\,$\mu$m band are detected with sufficient signal-to-noise ratios. 
The 3.3\,$\mu$m band has been enthusiastically studied for various astronomical objects, and is generally accepted as a fundamental C-H stretching band of astronomical PAHs, nevertheless less is known about the 3.4--3.6\,$\mu$m sub-features and the 5.25\,$\mu$m band. 
In particular, there are only a few detailed observational studies on the 5.25\,$\mu$m band due to its weakness and position, which is often located at the edge or gap of the detector spectral coverage. 
Here, the legacy of {\it AKARI}
gives us a great opportunity to investigate these minor bands together with the 3.3\,$\mu$m band 
based on a larger collection of spectra of Galactic H\,{I\hspace{-.1em}I} regions than ever before. 
In this section, we preset analysis results, 
and discuss the processing of carbonaceous grains in H\,{I\hspace{-.1em}I} regions. 

%%%%%%%%%%%%%%%%%%%%%%%%%%%%%%%%%%
\subsection{the 5.25\,$\mu$m band} \label{sec:discussion:1}
%%%%%%%%%%%%%%%%%%%%%%%%%%%%%%%%%%
Based on the results of their laboratory experiments and quantum-chemical calculations, 
previous studies indicate that 
the astronomical 5.25\,$\mu$m band could be a blend of overtone, difference, and combination bands of the fundamental C-H stretching and bending vibration modes \citep{alla89-b,boer09}. 
\cite{boer09} also suggest that C-C modes of large ionized PAHs contribute to those two bands. 
From the point of view of observations, \cite{boer09} report that the 5.25\,$\mu$m band strength is strongly correlated with the 11.3\,$\mu$m band strength, which supports their close connection with C-H vibration modes. 
However, their observational study is based on only four spectra obtained by the {\it ISO}/SWS for HD44179 (post-AGB star), NGC7027 (planetary nebula), and two positions in the Orion bar (H\,{I\hspace{-.1em}I} regions). 
A larger collection of astronomical spectra is important to make a firm conclusion. 

\begin{figure*}
\begin{center}
\includegraphics[scale=0.35]{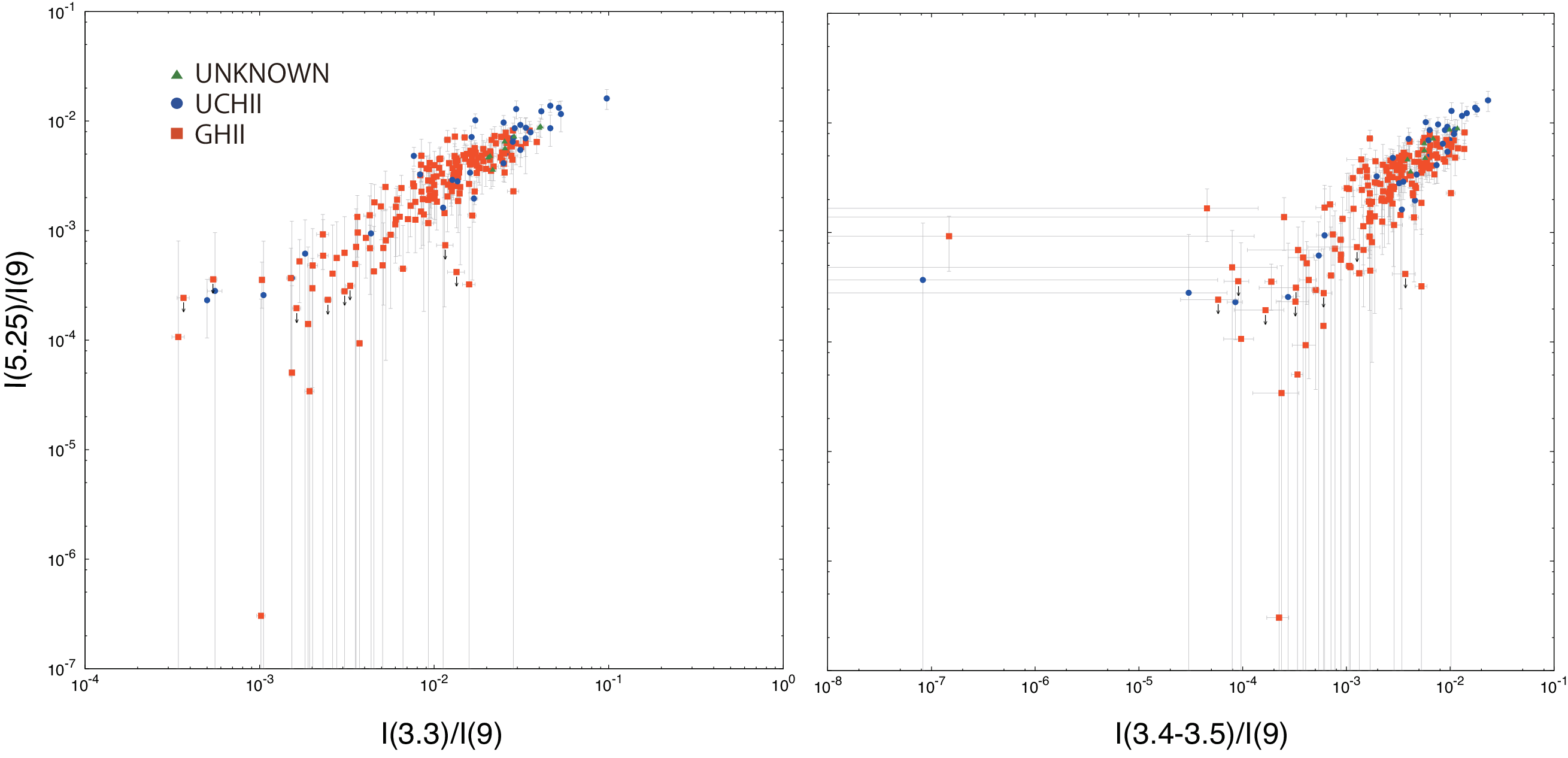}
\caption{
Plots of the ratio of the 5.25\,$\mu$m band to the {\it AKARI}  9\,$\mu$m surface brightness versus that of {\bf (a)} the 3.3\,$\mu$m band and {\bf (b)} the 3.4--3.6\,$\mu$m sub-features. 
The downward arrows indicate upper limits, where no apparent sign of the 5.25\,$\mu$m band is seen. 
}
\label{fig:4}
\end{center}
\end{figure*}

Here, in order to consider the nature of the 5.25\,$\mu$m band, 
we investigate their relative variation to the 3.3\,$\mu$m band and the 3.4--3.6\,$\mu$m sub-features (the summation of the 3.41 and 3.48\,$\mu$m components).  
Figures \ref{fig:4}{\bf a} and {\bf b} show the relative strength of the 5.25\,$\mu$m band with respect to the 9\,$\mu$m surface brightness against those of the 3.3\,$\mu$m band and the 3.4--3.6\,$\mu$m sub-features respectively. 
There is a tight correlation of the 5.25\,$\mu$m band both with the 3.3\,$\mu$m band and the 3.4--3.6\,$\mu$m sub-features 
(the weighted correlation coefficient $r$ is 0.92 and 0.88 respectively). 
As mentioned above, the 3.3\,$\mu$m band is assigned to an aromatic C-H stretching mode, 
and the 3.4--3.6\,$\mu$m sub-features to aliphatic C-H vibrational modes \citep[]{dule81}, 
or overtone of the fundamental aromatic C-H stretching mode. 
The present result confirms that the 5.25\,$\mu$m band has a close relation to C-H vibration modes. 
Given that the emission feature at a short wavelength like the 3.3\,$\mu$m band requires high excitation, it is compatible with the hypothesis that the overtone bands of fundamental C-H vibration modes contribute to the 5.25\,$\mu$m band. 
A slightly higher correlation is found with the 3.3\,$\mu$m band than with the 3.4--3.6\,$\mu$m sub-features, suggesting that the 5.25\,$\mu$m band has stronger connection to aromatic ones. 
Since the 3.4--3.6\,$\mu$m sub-features correlates with the 3.3\,$\mu$m band to some extent, 
the correlation between the 3.4--3.6\,$\mu$m sub-features and the 5.25\,$\mu$m band may be a secondary relation.  
However, the difference between those two correlations is not significant, and thus we cannot draw a clear conclusion on this point from the present data. 

We fit a linear function, $y = A_{\rm intercept} + B_{\rm slope} \cdot x$, 
where $y$ is the relative strength of the 5.25\,$\mu$m band to the 9\,$\mu$m surface brightness and $x$ is that of the 3.3\,$\mu$m band for all the data points and for those classified into UCH\,{I\hspace{-.1em}I} and GH\,{I\hspace{-.1em}I} regions separately, using the IDL routine ``FITEXY'' \citep{pres02}. 
Th fitting results are summarized in Table \ref{tab:5}.  
In all cases, the intercept $A_{\rm intercept}$ is very small, and 
the regression lines pass through the origin within uncertainty. 
The slope $B_{\rm slope}$ does not differ significantly between the UCH\,{I\hspace{-.1em}I} and H\,{I\hspace{-.1em}I} regions, and is in agreement with the intensity ratio of the 5.25\,$\mu$m band to the 3.3\,$\mu$m band of HD44179, NGC7027, and two positions in the Orion bar reported by \cite{boer09}, which ranges from 0.16 to 0.38 within a factor of two. 
These results suggest that the relative intensity of the 5.25\,$\mu$m band to the 3.3\,$\mu$m band is rather constant over a wide range of Galactic objects regardless their evolutionary stage.

%%%%%%%%%%%%%%%%%%%%%%%%%%%%%%%%%%
\subsection{the 3.4--3.6\,$\mu$m sub-features} \label{sec:discussion:2}
%%%%%%%%%%%%%%%%%%%%%%%%%%%%%%%%%%
It is widely known that the 3.3\,$\mu$m PAH band is commonly accompanied by adjacent satellite features at 3.41, 3.51, and 3.56\,$\mu$m and a broad plateau extending to 3.6\,$\mu$m for various kinds of astronomical objects \citep[e.g.][]{geba85,jobl96b}, 
yet the origin of these satellite features has not been clearly identified. 
Aliphatic hydrocarbon materials have a fundamental C-H stretching mode at 3.4\,$\mu$m \citep{pend02}. 
Hydrogenated PAHs also cause the 3.4\,$\mu$m band. 
Additional H atoms to peripheral C atoms of PAHs convert aromatic rings to aliphatic rings, which newly create aliphatic C-H stretching bands especially near 3.4\,$\mu$m and 3.5\,$\mu$m \citep{bern96}. 
Aliphatic side-groups at the periphery of PAHs like methyl (-CH$_3$), methylene (-CH$_2$-), and ethyl (-CH$_2$CH$_3$) also give rise to the satellite features around 3.4--3.5\,$\mu$m as aliphatic C-H stretching modes \citep[e.g.][]{dule81}. 
On the other hand, overtone bands of the aromatic C-H stretching mode also appear in that spectral range due to the anharmonicity \citep[at 3.41, 3.47, 3.5, and 3.56\,$\mu$m,][]{bark87}. 
\cite{jobl96b} report the 3\,$\mu$m spectrum variation within the reflection nebulae NGC~1333 SVS3 and NGC~2023, suggesting that the observed 3.4\,$\mu$m band is too intense to originate from overtone bands of the aromatic C-H stretching mode alone, and that the relative variation of the 3.4\,$\mu$m band with respect to the 3.3\,$\mu$m band can be explained by  photochemical erosion of aliphatic side-groups attached to PAHs due to the strong FUV radiation field. 

\begin{figure*}
\begin{center}
\includegraphics[scale=0.6]{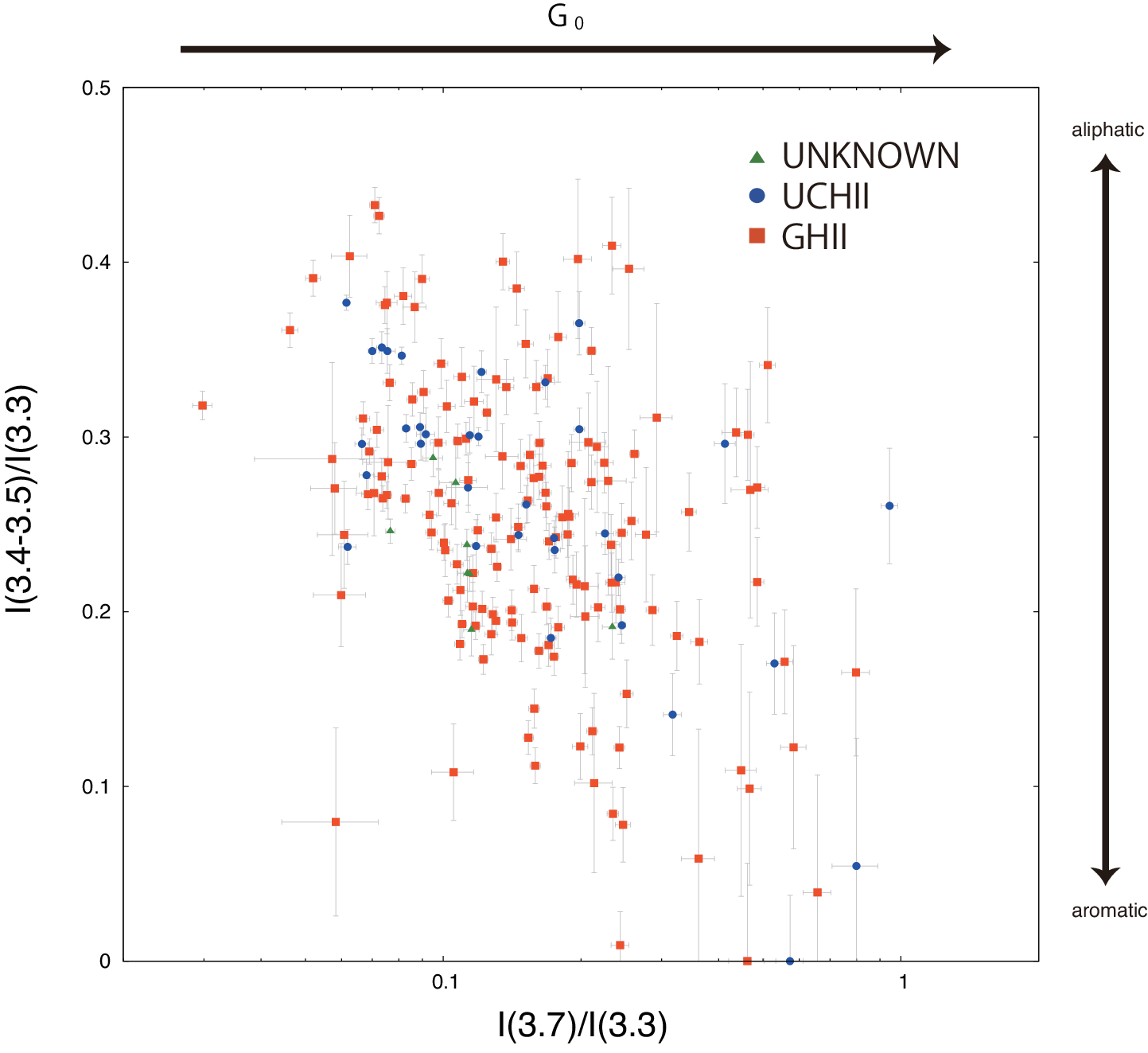}
\caption{
The diagram of the relative intensity ratio of the 3.4--3.6\,$\mu$m sub-features  to the 3.3\,$\mu$m band, 
$I_{\rm 3.4-3.6\,\mu m}/I_{\rm 3.3\,\mu m}$, 
versus that of the 3.7\,$\mu$m continuum intensity to the 3.3\,$\mu$m band, $I_{\rm cont, 3.7\,\mu m}/I_{\rm 3.3\,\mu m}$}. 
\label{fig:5}
\end{center}
\end{figure*}

In spectra of reflection nebulae, 
\cite{sell84} indicate the presence of the continuum emission with a high color temperature of $\sim$1000\,K in the NIR range.  
From observations of the Orion Nebulae with the Wide Field Cryogenic Telescope-{I\hspace{-.1em}I}, 
\cite{hara12} suggest that 
after subtraction of the free-free emission from the 3.7\,$\mu$m continuum  
there remains significant residual component, 
and that its relative strength to the 3.3\,$\mu$m band emission reflects the ionization degree of PAHs. 
In the present study, 
the relative intensity of the 3.4--3.6\,$\mu$m sub-features to the 3.3\,$\mu$m band, 
$I_{\rm 3.4-3.6\,\mu m}/I_{\rm 3.3\,\mu m}$, 
lies in a range of relatively small values, 0.0--0.5, 
which implies that the aliphatic structure is a small part of the band carriers 
at least for the present target sources \citep[e.g.][]{lidr12}, 
but exhibits a small, but clear systematic variation 
against the intensity ratio of the 3.7\,$\mu$m continuum intensity to the 3.3\,$\mu$m band, 
$I_{\rm cont, 3.7\,\mu m}/I_{\rm 3.3\,\mu m}$ (see Figure \ref{fig:5}). 
The 3.7\,$\mu$m continuum intensity, $I_{\rm cont, 3.7\,\mu m}$, is the integrated flux over a 3.65--3.71\,$\mu$m range. 
Hydrogen free-free emission and free-bound and helium free-free emission are estimated from the intensity of Br$\alpha$
and are subtracted from the observed continuum. 
As suggested by \cite{hara12}, 
we assume that the ratio of $I_{\rm cont, 3.7\,\mu m}/I_{\rm 3.3\,\mu m}$ mirrors the ionization degree of PAHs. 
As the ionization degree of PAHs increases, 
the ratio of $I_{\rm 3.4-3.6\,\mu m}/I_{\rm 3.3\,\mu m}$ slightly decreases with a large scatter. 

The aliphatic fraction of the band emitters is thought to be a major factor to influence the observed intensity ratio of $I_{\rm 3.4-3.6\,\mu m}/I_{\rm 3.3\,\mu m}$. 
Aliphatic C-H bonds are less resilient than aromatic ones and they need less energy to break. 
The laboratory experiments reveal that 
irradiation of UV light and heating drive hydrocarbon solid materials to evolve to aromatic-rich materials 
\citep[e.g.][]{iida85,smit84,saka90}. 
Recently \cite{jone12-a,jone12-b,jone12-c} presents a theoretical model for the evolution of amorphous hydrocarbon materials, 
and suggests that UV-photolysis leads to the ultimate transition towards aromatic-dominated materials 
especially for small grains 
($a${\footnotesize \hspace{0.3em}\raisebox{0.4ex}{$<$}\hspace{-0.75em}\raisebox{-.7ex}{$\sim$}\hspace{0.3em}}20\,nm). 
Given that the grains in this size range dominate the 3.3--3.6\,$\mu$m band emission \citep[e.g.][]{schu93,drli07}, 
his prediction can be applicable to our study. 
The PAH ionization degree is controlled by a balance between photo-ionization and recombination with ambient electrons. 
Among the present target regions, 
the variation of the UV radiation field is expected to be the most contributing factor in the difference of the PAH ionization degree.  
The observed ratio of $I_{\rm 3.4-3.6\,\mu m}/I_{\rm 3.3\,\mu m}$ decreases with the increase of the PAH ionization degree. 
This trend is consistent with the scenario that aliphatic C-H bonds in the band emitters are efficiently destroyed prior to aromatic ones in ionized gas and/or its boundary with the evolution of the UV radiation field. 
If a photo-thermal process is the main cause of the decrease of the 3.4--3.6\,$\mu$m sub-features relative to the 3.3\,$\mu$m band, 
we can investigate the evolutionary history of hydrocarbon materials in interstellar environments 
from the intensity ratio of $I_{\rm 3.4-3.6\,\mu m}/I_{\rm 3.3\,\mu m}$. 
\cite{kane12} and \cite{yama12} report 
the enhancement of the 3.4--3.6\,$\mu$m sub-features relative to the 3.3\,$\mu$m band 
in the CO molecular loop near the Galactic center and in the halo of M82. 
They conclude that this enhancement arises from shattering of carbonaceous grains driven by shock winds \citep{jone96}. 
In contrast, the present study indicates no such enhancement in the ionized gas-dominated regions. 
Therefore, it can be inferred that shattering is not an effective process in the interior and the periphery of H\,{I\hspace{-.1em}I} regions, which is consistent with the model prediction of \cite{mice10b}. 
Some previous studies suggest that the variation of the 3.4--3.6\,$\mu$m sub-features relative to the 3.3\,$\mu$m band reflects the processing of interstellar hydrocarbon materials due to thermal annealing, 
but they are discussed only for a limited number of astronomical objects 
including H\,{I\hspace{-.1em}I} regions, planetary nebulae, and reflection nebulae (e.g. \citealt{jobl96b}; \citealt{goto03,boul11}). 
The present study is the first investigation 
based on a large quantity of high-quality spectra of various Galactic H\,{I\hspace{-.1em}I} regions, 
and gives the first clear evidence for the dust processing in ionized medium.

%%%%%%%%%%%%%%%%%%%%%%%%%%%%%%%%%%
\subsection{Variation of the MIR color} \label{sec:discussion:3}
%%%%%%%%%%%%%%%%%%%%%%%%%%%%%%%%%%
The {\it AKARI} {\it S9W} band filter covers a wavelength range of 6.7--11.6\,$\mu$m, 
which includes the prominent PAH emission band features at 6.2, 7.7, 8.6, and 11.3\,$\mu$m.  
Therefore, it can be conjectured that the 9\,$\mu$m band intensity mostly originates in these PAH emission features \citep[e.g.][]{ishi07,kane12}. 
On the other hand, the origin of the 18\,$\mu$m emission is still ambiguous. 
In the report of {\it AKARI} imaging observations of the reflection nebulae IC~4954 and IC~4955 region, 
\cite{ishi07} suggest that the continuum emission originating from stochastic heating of very small grains (VSGs) dominates spectra in the wavelength range of the {\it L18W} band filter. 
However, this assumption does not always hold good in the case of H\,{I\hspace{-.1em}I} regions, the target objects of this study. 
In H\,{I\hspace{-.1em}I} regions, the intensity of radiation field reaches 10$^4$--10$^5$ times of that in the solar vicinity  
\citep[e.g.][]{tiel93,bern09,salg12}. 
In such a harsh environment, the equilibrium temperature of big grains (BGs) becomes higher 
and thermal emission can be dominant even at short wavelengths around 18\,$\mu$m. 

\begin{figure*}
\begin{center}
\includegraphics[scale=0.6]{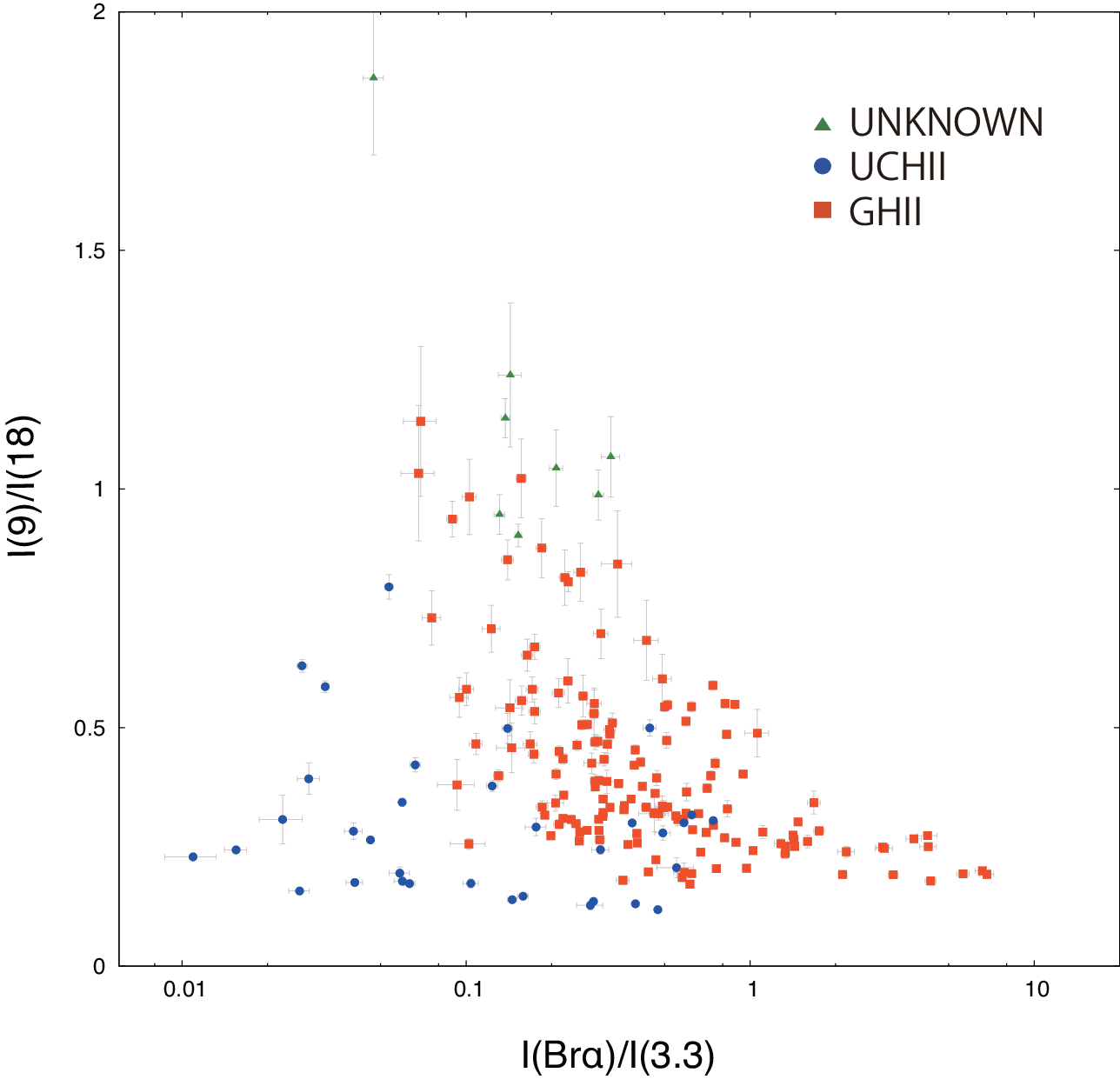}
\caption{
The diagram of the MIR color of the {\it AKARI} 9\,$\mu$m to 18\,$\mu$m band, 
$I_{\rm 9\,\mu m}/I_{\rm 18\,\mu m}$, 
against the relative intensity of the hydrogen recombination line Br$\alpha$ at 4.05\,$\mu$m to the 3.3\,$\mu$m band, 
$I_{\rm Br\alpha}/I_{\rm 3.3\,\mu m}$. }
\label{fig:6}
\end{center}
\end{figure*}

As shown in Figure \ref{fig:6}, 
the MIR color of {\it AKARI} 9\,$\mu$m to 18\,$\mu$m band steeply declines 
against the ratio of $I_{\rm Br\alpha}/I_{\rm 3.3\,\mu m}$. 
The ratio of $I_{\rm Br\alpha}/I_{\rm 3.3\,\mu m}$ is considered as a good indicator of the fraction of the ionized gas along the line of sight. 
The present result suggests that the MIR excess around 18\,$\mu$m becomes stronger relative to the PAH 9\,$\mu$m emission 
with the transition from PDRs to ionized-gas dominated regions. 
The observed trend can be explained by PAH destruction and either VSG or BG replenishment inside the ionized medium, 
which are proposed by several authors as a possible interpretation of Herschel and Spitzer multi-wavelenght observations \citep[e.g.][]{para11,pala12}. 
On the other hand, 
it is expected that the temperature of BGs is raised with the increase of incident radiation field from PDRs to ionized-gas, 
which can also contribute to the observed trend. 
The present result may reflect the variation in the incident radiation field between PDRs and ionized gas rather than that of dust abundance. 
These two effects cannot be distinguished from the present dataset and 
we cannot draw a clear conclusion at the moment. 
In comparison with the GH\,{I\hspace{-.1em}I} regions, 
the UCH\,{I\hspace{-.1em}I} regions are distributed over a different region in Figure \ref{fig:6}, 
in the relatively left side of the diagram. 
This might result from the different geometric structures between them, that is, 
larger contamination from their surroundings in the UCH\,{I\hspace{-.1em}I} sources. 
{\it AKARI} MIR survey covers 99\% of the whole sky. 
The wide scope of the {\it AKARI} MIR survey will be very helpful to investigate the dust processing in the ISM.

%%%%%%%%%%%%%%%%%%%%%%%%%%%%%%%%%%
%%%%%%%%%%%%%%%%%%%%%%%%%%%%%%%%%%
\section{Summary and Conclusion} \label{sec:summary}
%%%%%%%%%%%%%%%%%%%%%%%%%%%%%%%%%%
%%%%%%%%%%%%%%%%%%%%%%%%%%%%%%%%%%
We analyze the NIR spectra of the diffuse emission lights 
of 36 Galactic H\,{I\hspace{-.1em}I} region or H\,{I\hspace{-.1em}I} region-like objects
with the IRC onboard {\it AKARI}. 
The targets include 7 UCH\,{I\hspace{-.1em}I} and 27 GH\,{I\hspace{-.1em}I} regions 
plus two H\,{I\hspace{-.1em}I} region-like infrared sources. 
By virtue of the special calibration mode, 
we are able to extract spectra taken with 
the grism and prism mode (2.5--5.0\,$\mu$m, $R$\,$\sim$\,100 and 1.7--5.4\,$\mu$m, $R$\,$\sim$\,20--40)
at the same place on the sky. 

The spectra exhibit a variety of ISM features 
such as helium and hydrogen recombination lines and 
the PAH emission features including the 5.25\,$\mu$m band, 
which is present in the unique wavelength coverage of the prism spectroscopic mode. 
Some spectra also show the clear absorption features 
of H$_2$O and CO ices at 3.05\,$\mu$m and 4.27\,$\mu$m. 

The 5.25\,$\mu$m band strongly correlates with both the 3.3\,$\mu$m band and the 3.4--3.6\,$\mu$m sub-features, 
which proves its close relationship to the C-H vibration modes of astronomical PAHs. 
The relative intensity of the 5.25\,$\mu$m band to the 3.3\,$\mu$m band does not vary significantly 
between different kinds of astronomical objects in our Galaxy. 
Combined with the PAH emission features involving C-H vibrations at longer wavelengths such as the 11.3\,$\mu$m band,
the 5.25\,$\mu$m band could be a good indicator of PAH size distribution as the 3.3\,$\mu$m band 
\citep[e.g.][]{schu93,mori12}. 
 
The observed intensity ratio of the 3.4--3.6\,$\mu$m sub-features to the 3.3$\mu$m band decreases 
as the intensity ratio of the 3.7\,$\mu$m continuum to the 3.3\,$\mu$m band increases. 
This result is the first clear evidence for erosion of the aliphatic structure of the band carriers inside or at the boundary of the ionized domain. 
We also report that the 9\,$\mu$m to 18\,$\mu$m color falls rapidly against the ratio of $I_{\rm Br\alpha}/I_{\rm 3.3\,\mu m}$. 
The result does not contradict the model of PAH destruction and VSG or BG reproduction inside the ionized medium, 
but other interpretations cannot be ruled out. 

The present study demonstrates the wealth of the {\it AKARI}/IRC NIR spectroscopy for the understanding of the ISM physics, 
but covers only a small part of the legacy of {\it AKARI}. 
The further investigation on other ISM features, 
and with a much more larger sample will be reported separately. 
The {\it AKARI}/IRC spectroscopic data archives are now in progress. 
We will soon release the spectral catalogue including the present reduced data.

%%%%%%%%%%%%%%%%%%%%%%%%%%%%%%%%%%
%%%%%%%%%%%%%%%%%%%%%%%%%%%%%%%%%%
\section*{Acknowledgements}
%%%%%%%%%%%%%%%%%%%%%%%%%%%%%%%%%%
%%%%%%%%%%%%%%%%%%%%%%%%%%%%%%%%%%
We would like to thank the anonymous referee for his/her useful comments, which improved the paper significantly. 
This work is based on observations with {\it AKARI}, 
a Japan Aerospace Exploration Agency (JAXA) project with the participation of European Space Agency (ESA). 
We would like to express our heartfelt gratitude to all the members of the {\it AKARI} project 
for their long-term support and constant encouragement. 
The laboratory data of the H$_2$O ice were obtained from the Leiden atomic and molecular database. 
This work is supported in part by 
a Grant-in-Aid for Scientific Research from the Japan Society of Promotion of Science (JSPS). 
T.~I.~M. and R.~O. receive financial support from a Grant-in-Aid for JSPS Fellows.

%%%%%%%%%%%%%%%%%%%%%%%%%%%%%%%%%%
%%%%%%%%%%%%%%%%%%%%%%%%%%%%%%%%%%
\appendix
\section*{Appendix A. Comparison of the grism and prism spectra} \label{app:A}
\begin{figure*}
\begin{center}
\includegraphics[scale=0.6]{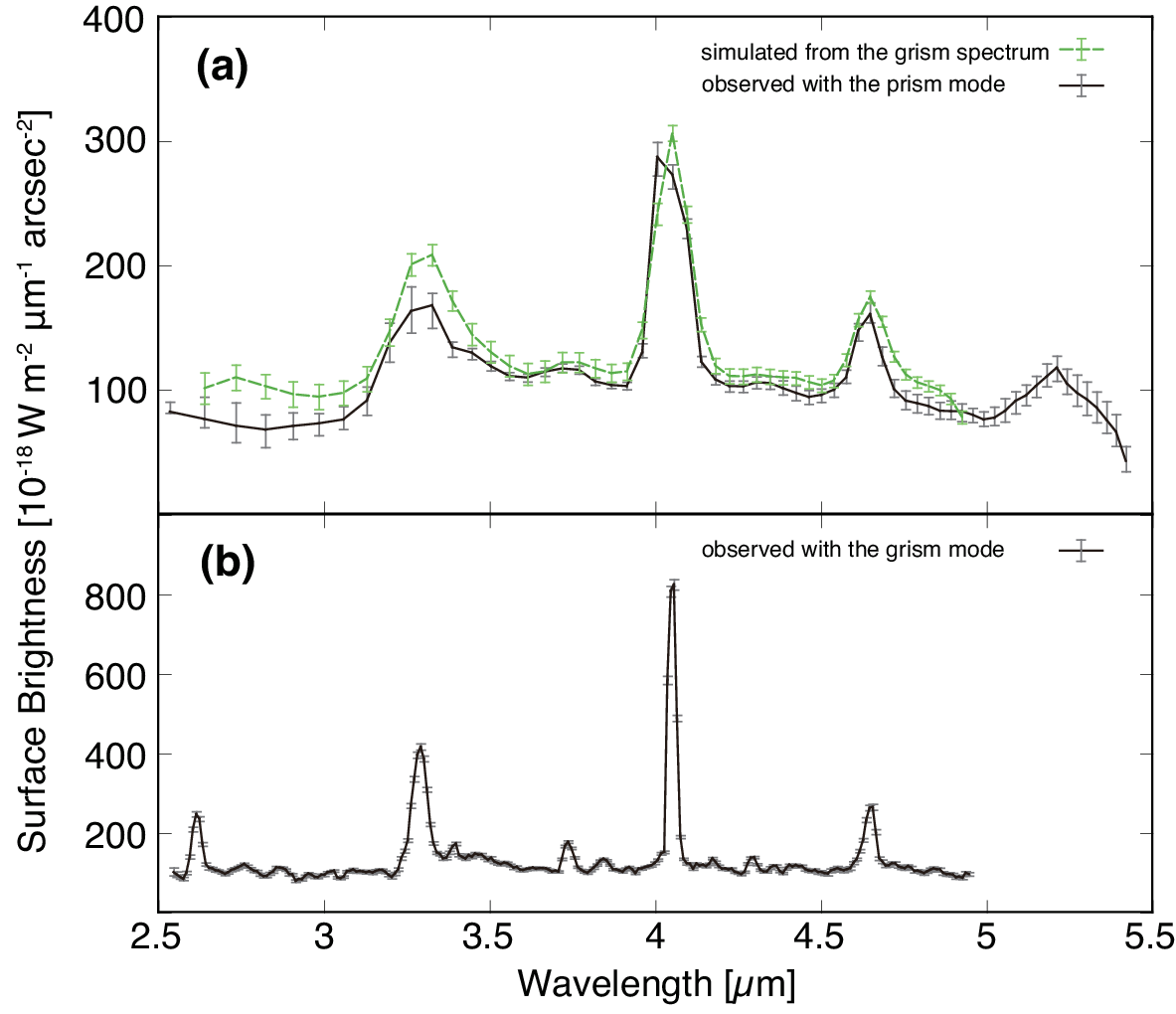}
\caption{
{\bf (a)} Example of the comparison of the prism spectrum (black solid line) and the simulated spectrum (green dashed line). 
The corresponding grism spectrum is shown in the lower panel {\bf(b)}. 
The spectra are those taken towards W49A (ID:\,5200299.1). 
}
\label{fig:A1}
\end{center}
\end{figure*}

To make quantitative comparison of the spectra taken with the grism mode and the prism mode, 
we developed software to simulate a spectrum of the prism mode from a spectrum taken with the grism mode.
In the simulation, we take into account that the spectral dispersion in the prism mode changes with the wavelength. 
We disperse the photons detected in a spectral element in the grism spectrum into a spectral element of the prism spectrum taking account of the slit width. 
While the slit is located at the same position on the sky, 
the location on the array differs between the grism and prism modes due to the different optical alignment. 
We correct for the position shift along the slit direction between the grism and prism modes 
and extract the spectra at the same position (see also \S \ref{sec:observations:2}). 
To minimize a possible shift remaining, we compare spectra of regions that do not show particular spatial structures. 
Figure \ref{fig:A1}{\bf a} shows the prism and simulated spectra of W49A (ID: 5200299.1) as an example of the comparison 
and Figure \ref{fig:A1}{\bf b} plots the corresponding spectrum taken with the grism mode for reference.
The simulated spectrum shown by the dashed line is in good agreement with the spectrum taken with the prism mode (solid line). 
Note that the grism spectrum does not provide the data needed for the both ends of the simulated spectrum 
and thus the deviations at the ends come partly from the insufficient data for the simulation. 
The band emission in the simulated spectrum reproduces that seen in the prism spectrum fairly well,
confirming that the relative calibration between the prism and grism spectra is within the given uncertainty 
and assuring that the software simulates the prism spectrum reliably. 
There is a small systematic difference in the continuum level 
and the simulated spectrum is slightly brighter than the actual prism spectrum particularly around 4.7--4.9\,$\mu$m.   
It may come from the residual in the dark current correction, 
but does not affect the band profile. 

\begin{figure*}
\begin{center}
\includegraphics[scale=0.6]{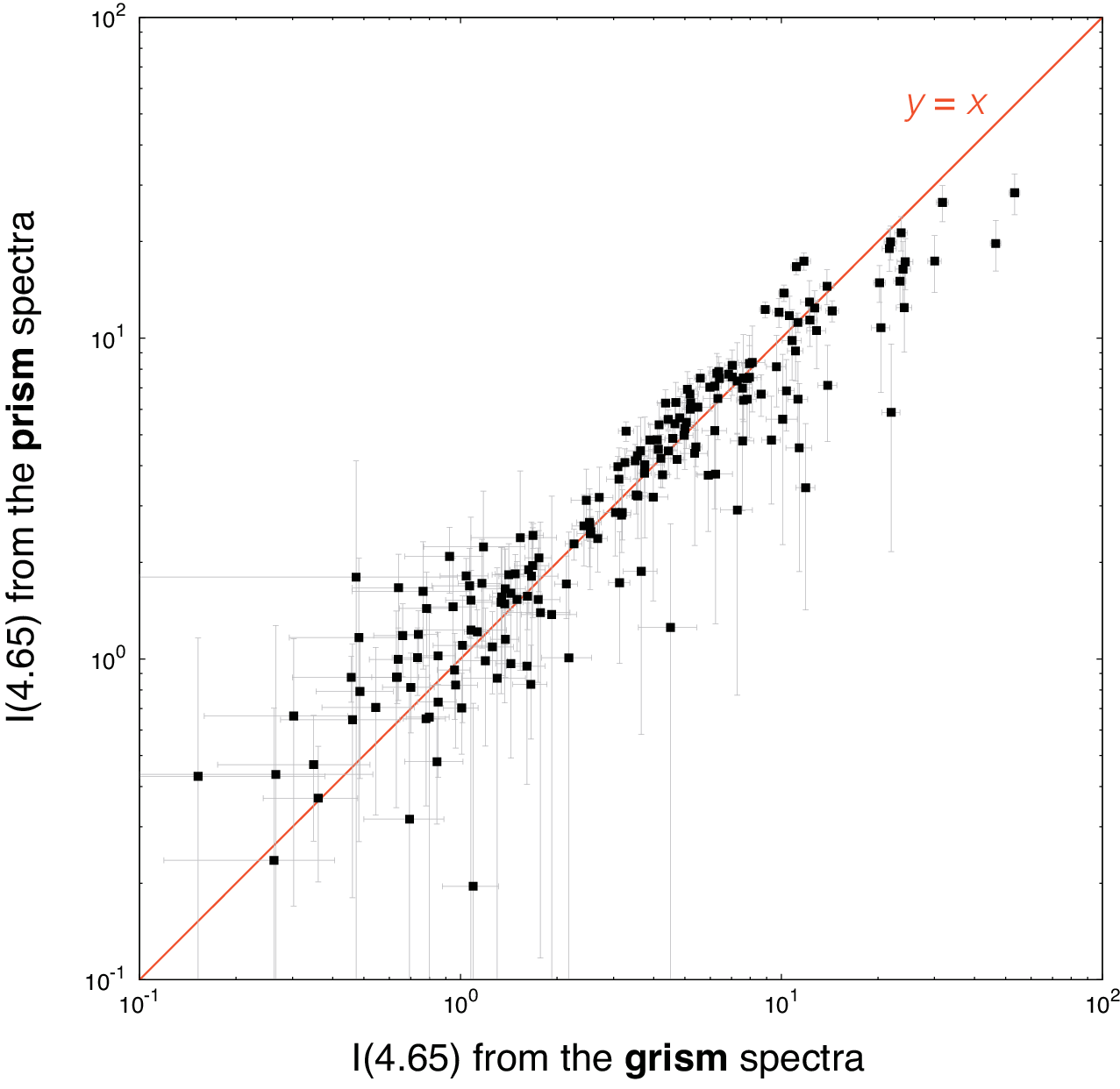}
\caption{
Plot of the intensity of the hydrogen recombination line Pf$\beta$ at 4.65\,$\mu$m 
measured from the prism spectra by the spectral fitting described in \S \ref{sec:analysis:2}, 
against that of the grism spectra. 
We exclude those which exhibit a serious effect of saturation around the peak of Pf$\beta$ from the plot. 
}
\label{fig:A2}
\end{center}
\end{figure*}

Based upon the above result, 
we fit the prism spectra 
with the simulated spectral profiles of each component (see details in \S \ref{sec:analysis:2}). 
By this technique, we estimate the intensity of the hydrogen recombination line Pf$\beta$ at 4.65\,$\mu$m from the prism spectra, 
which is in good agreement with the results of the spectral fitting to the grism spectra (see Figure \ref{fig:A2}). 
This fact establishes the validity of our fitting method, 
and also confirms the accuracy of the absolute flux calibration of the {\it AKARI}/IRC at NIR wavelengths. 

%\section*{Appendix B. Gallery of the grism and prism spectra and the target S9W Images} \label{app:B} 
%We summarize the NIR spectra and the {\it S9W} image of each target object in Figure \ref{fig:A3}. 

%%%%%%%%%%%%%%%%%%%%%%%%%%%%%%%%%%
%%%%%%%%%%%%%%%%%%%%%%%%%%%%%%%%%%

%%%%%%%%%%%%%%%%%%%%%%%%%%%%%%%%%%
%%%%%%%%%%%%%%%%%%%%%%%%%%%%%%%%%%
\bibliographystyle{apj}
\bibliography{reference}
%%%%%%%%%%%%%%%%%%%%%%%%%%%%%%%%%%
%%%%%%%%%%%%%%%%%%%%%%%%%%%%%%%%%%

%%%%%%%%%%%%%%%%%%%%%%%%%%%%%%%%%%
%%%%%%%%%%%%%%%%%%%%%%%%%%%%%%%%%%
\tabletypesize{\scriptsize}
%ttttttttttttttttttttttttttttttttttttttttttttttttttttttttttttttttttttttttt%
\begin{deluxetable}{lccccccccc} 
\tablecolumns{10} 
\tablewidth{0pt} 
\tablecaption{Observation log and target parameters\label{tab:1}}
\tablehead{
       \multicolumn{2}{c}{Target}    
    & \colhead{$d$}    
    & \colhead{log{\it N}}  
    & \multicolumn{2}{c}{Center position of the slit}   
    & \colhead{Obs.Date}  
    & \colhead{Obs.ID}    
    & \colhead{Obs.AOT\tablenotemark{\,{\dag_2}}}    
    & \colhead{Disperser}\\
       \colhead{Name} 
    & \colhead{Type\tablenotemark{\,{\dag_1}}} 
    & \colhead{\,\,\,\,\,\,\,\,\,[kpc]\tablenotemark{\,{\dag_1}}} 
    & \colhead{\,\,\,\,\,\,\,\,\,\,LyC\tablenotemark{\,{\dag_1}}} 
    & \colhead{$l$} 
    & \colhead{$b$} 
    & 
    & 
    & 
    &}
\startdata 
M8              & GH\,{I\hspace{-.1em}I}  &  2.8 & 50.19   &     5.972 &    -1.172 & 2008-09-24 & 5200162.1 &  IRCZ4 c;Ns & NG and NP \\
M8              & GH\,{I\hspace{-.1em}I}  &  2.8 & 50.19   &     5.974 &    -1.176 & 2008-09-23 & 5200161.1 &  IRCZ4 c;Nh & NG and NP \\
G8.137+0.228    & GH\,{I\hspace{-.1em}I}  & 13.5 & 50.47   &     8.142 &     0.225 & 2008-09-23 & 5200164.1 &  IRCZ4 c;Ns & NG and NP \\
G8.137+0.228    & GH\,{I\hspace{-.1em}I}  & 13.5 & 50.47   &     8.143 &     0.223 & 2008-09-22 & 5200163.1 &  IRCZ4 c;Nh & NG and NP \\
W31a            & GH\,{I\hspace{-.1em}I}  &  4.5 & 50.66   &    10.163 &    -0.358 & 2008-09-25 & 5200166.1 &  IRCZ4 c;Ns & NG and NP \\
W31a            & GH\,{I\hspace{-.1em}I}  &  4.5 & 50.66   &    10.165 &    -0.362 & 2008-09-24 & 5200165.1 &  IRCZ4 c;Nh & NG and NP \\
W31b            & GH\,{I\hspace{-.1em}I}  & 15.0 & 50.90   &    10.324 &    -0.158 & 2008-09-24 & 5200167.1 &  IRCZ4 c;Nh & NG and NP \\
M17b            & GH\,{I\hspace{-.1em}I}  &  2.4 & 51.22   &    15.027 &    -0.693 & 2008-09-28 & 5200172.1 &  IRCZ4 c;Ns & NG and NP \\
M17b            & GH\,{I\hspace{-.1em}I}  &  2.4 & 51.22   &    15.031 &    -0.696 & 2008-09-28 & 5200171.1 &  IRCZ4 c;Nh & NG and NP \\
M17a            & GH\,{I\hspace{-.1em}I}  &  2.4 & 51.22   &    15.048 &    -0.677 & 2008-09-27 & 5200169.1 &  IRCZ4 c;Nh & NG and NP \\
W42             & GH\,{I\hspace{-.1em}I}  & 11.5 & 50.93   &    25.384 &    -0.179 & 2008-10-02 & 5200294.2 &  IRCZ4 c;Nh & NG and NP \\
W42             & GH\,{I\hspace{-.1em}I}  & 11.5 & 50.93   &    25.385 &    -0.180 & 2008-10-02 & 5200294.1 &  IRCZ4 c;Nh & NG and NP \\
G29.944-0.042   & GH\,{I\hspace{-.1em}I}  &  6.2 & 50.33   &    29.958 &    -0.017 & 2008-10-04 & 5200295.2 &  IRCZ4 c;Nh & NG and NP \\
G29.944-0.042   & GH\,{I\hspace{-.1em}I}  &  6.2 & 50.33   &    29.960 &    -0.017 & 2008-10-04 & 5200295.1 &  IRCZ4 c;Nh & NG and NP \\
W49A            & GH\,{I\hspace{-.1em}I}  & 11.8 & 51.21   &    43.175 &     0.004 & 2008-10-13 & 5200299.2 &  IRCZ4 c;Nh & NG and NP \\
W49A            & GH\,{I\hspace{-.1em}I}  & 11.8 & 51.21   &    43.176 &     0.003 & 2008-10-12 & 5200299.1 &  IRCZ4 c;Nh & NG and NP \\
G48.596+0.042   & GH\,{I\hspace{-.1em}I}  &  9.8 & 50.14   &    48.610 &     0.027 & 2008-10-17 & 5200300.1 &  IRCZ4 c;Nh & NG and NP \\
G48.596+0.042   & GH\,{I\hspace{-.1em}I}  &  9.8 & 50.14   &    48.611 &     0.028 & 2008-10-17 & 5200300.2 &  IRCZ4 c;Nh & NG and NP \\
W51             & GH\,{I\hspace{-.1em}I}  &  5.5 & 50.03   &    48.915 &    -0.286 & 2008-10-18 & 5200301.2 &  IRCZ4 c;Nh & NG and NP \\
W51             & GH\,{I\hspace{-.1em}I}  &  5.5 & 50.03   &    48.916 &    -0.286 & 2008-10-17 & 5200301.1 &  IRCZ4 c;Nh & NG and NP \\
W58A            & GH\,{I\hspace{-.1em}I}  &  8.6 & 50.06   &    70.289 &     1.602 & 2009-05-02 & 5200767.1 &  IRCZ4 c;Nh & NG and NP \\
G70.293+1.600   & UCH\,{I\hspace{-.1em}I} &  8.6 & 49.29   &    70.295 &     1.602 & 2008-11-06 & 5200337.1 &  IRCZ4 c;Nh & NG and NP \\
W58A            & GH\,{I\hspace{-.1em}I}  &  8.6 & 50.06   &    70.295 &     1.602 & 2009-11-06 & 5201198.1 &  IRCZ4 c;Nh & NG and NP \\
G75.783+0.343   & UCH\,{I\hspace{-.1em}I} &  4.1 & 46.78   &    75.765 &     0.342 & 2009-05-11 & 5200772.1 &  IRCZ4 c;Nh & NG and NP \\
G76.383-0.621   & UCH\,{I\hspace{-.1em}I} &  1.0 & 45.06   &    76.379 &    -0.619 & 2008-11-15 & 5200344.1 &  IRCZ4 c;Ns & NG and NP \\
G76.383-0.621   & UCH\,{I\hspace{-.1em}I} &  1.0 & 45.06   &    76.385 &    -0.622 & 2008-11-15 & 5200343.1 &  IRCZ4 c;Nh & NG and NP \\
G78.438+2.659   & UCH\,{I\hspace{-.1em}I} &  3.3 & 46.83   &    78.437 &     2.659 & 2009-05-13 & 5200776.1 &  IRCZ4 c;Nh & NG and NP \\
G78.438+2.659   & UCH\,{I\hspace{-.1em}I} &  3.3 & 46.83   &    78.438 &     2.657 & 2009-05-13 & 5200777.1 &  IRCZ4 c;Ns & NG and NP \\
DR7             & GH\,{I\hspace{-.1em}I}  &  8.3 & 50.15   &    79.302 &     1.305 & 2009-05-17 & 5200769.1 &  IRCZ4 c;Nh & NG and NP \\
DR7             & GH\,{I\hspace{-.1em}I}  &  8.3 & 50.15   &    79.304 &     1.304 & 2009-05-18 & 5200770.1 &  IRCZ4 c;Ns & NG and NP \\
G81.679+0.537   & UCH\,{I\hspace{-.1em}I} &  2.0 & 47.52   &    81.682 &     0.542 & 2008-11-22 & 5200347.1 &  IRCZ4 c;Nh & NG and NP \\
G111.282-0.663  & UCH\,{I\hspace{-.1em}I} &  2.5 & 46.68   &   111.282 &    -0.661 & 2009-01-18 & 5200433.1 &  IRCZ4 c;Ns & NG and NP \\
G111.282-0.663  & UCH\,{I\hspace{-.1em}I} &  2.5 & 46.68   &   111.286 &    -0.660 & 2009-01-16 & 5200432.1 &  IRCZ4 c;Nh & NG and NP \\
G133.947+1.064  & UCH\,{I\hspace{-.1em}I} &  3.0 & 47.92   &   133.950 &     1.064 & 2009-08-20 & 5200959.1 &  IRCZ4 c;Ns & NG and NP \\
RCW42           & GH\,{I\hspace{-.1em}I}  &  6.4 & 50.36   &   274.004 &    -1.146 & 2008-12-15 & 5200452.1 &  IRCZ4 c;Nh & NG and NP \\
RCW42           & GH\,{I\hspace{-.1em}I}  &  6.4 & 50.36   &   274.008 &    -1.146 & 2008-12-15 & 5200453.1 &  IRCZ4 c;Ns & NG and NP \\
G282.023-1.180  & GH\,{I\hspace{-.1em}I}  &  5.9 & 50.32   &   282.022 &    -1.182 & 2009-01-01 & 5200436.1 &  IRCZ4 c;Nh & NG and NP \\
G282.023-1.180  & GH\,{I\hspace{-.1em}I}  &  5.9 & 50.32   &   282.025 &    -1.183 & 2009-01-02 & 5200437.1 &  IRCZ4 c;Ns & NG and NP \\
RCW49           & GH\,{I\hspace{-.1em}I}  &  4.7 & 50.96   &   284.299 &    -0.346 & 2009-01-04 & 5200438.1 &  IRCZ4 c;Nh & NG and NP \\
RCW49           & GH\,{I\hspace{-.1em}I}  &  4.7 & 50.96   &   284.303 &    -0.347 & 2009-01-04 & 5200439.1 &  IRCZ4 c;Ns & NG and NP \\
NGC3372         & GH\,{I\hspace{-.1em}I}  &  2.5 & 50.11   &   287.377 &    -0.628 & 2009-01-10 & 5200440.1 &  IRCZ4 c;Nh & NG and NP \\
NGC3372         & GH\,{I\hspace{-.1em}I}  &  2.5 & 50.11   &   287.381 &    -0.631 & 2009-01-10 & 5200441.1 &  IRCZ4 c;Ns & NG and NP \\
G289.066-0.357  & GH\,{I\hspace{-.1em}I}  &  7.9 & 50.05   &   289.064 &    -0.358 & 2009-01-13 & 5200442.1 &  IRCZ4 c;Nh & NG and NP \\
G289.066-0.357  & GH\,{I\hspace{-.1em}I}  &  7.9 & 50.05   &   289.068 &    -0.359 & 2009-01-13 & 5200443.1 &  IRCZ4 c;Ns & NG and NP \\
NGC3576         & GH\,{I\hspace{-.1em}I}  &  3.1 & 50.28   &   291.282 &    -0.713 & 2009-01-17 & 5200444.1 &  IRCZ4 c;Nh & NG and NP \\
NGC3576         & GH\,{I\hspace{-.1em}I}  &  3.1 & 50.28   &   291.286 &    -0.715 & 2009-01-17 & 5200445.1 &  IRCZ4 c;Ns & NG and NP \\
NGC3603         & GH\,{I\hspace{-.1em}I}  &  7.9 & 51.50   &   291.608 &    -0.529 & 2009-01-17 & 5200446.1 &  IRCZ4 c;Nh & NG and NP \\
NGC3603         & GH\,{I\hspace{-.1em}I}  &  7.9 & 51.50   &   291.611 &    -0.530 & 2009-01-17 & 5200447.1 &  IRCZ4 c;Ns & NG and NP \\
G305.359+0.194  & GH\,{I\hspace{-.1em}I}  &  3.5 & 50.13   &   305.353 &     0.197 & 2009-08-09 & 5200932.2 &  IRCZ4 c;Ns & NG and NP \\
G305.359+0.194  & GH\,{I\hspace{-.1em}I}  &  3.5 & 50.13   &   305.354 &     0.196 & 2009-08-09 & 5200932.1 &  IRCZ4 c;Ns & NG and NP \\
G319.158-0.398  & GH\,{I\hspace{-.1em}I}  & 11.5 & 50.30   &   319.167 &    -0.422 & 2009-08-25 & 5200933.1 &  IRCZ4 c;Nh & NG and NP \\
G319.392-0.009  & GH\,{I\hspace{-.1em}I}  & 11.5 & 50.17   &   319.400 &    -0.011 & 2009-08-26 & 5200936.2 &  IRCZ4 c;Ns & NG and NP \\
G319.392-0.009  & GH\,{I\hspace{-.1em}I}  & 11.5 & 50.17   &   319.402 &    -0.011 & 2009-08-26 & 5200936.1 &  IRCZ4 c;Ns & NG and NP \\
G330.868-0.365  & GH\,{I\hspace{-.1em}I}  & 10.8 & 50.56   &   330.888 &    -0.372 & 2008-09-02 & 5200109.1 &  IRCZ4 c;Nh & NG and NP \\
G331.386-0.359  & UNKNOWN  &  -- &  --   &   331.390 &    -0.360 & 2008-09-03 & 5200113.1 &  IRCZ4 c;Nh & NG and NP \\
G333.122-0.446  & GH\,{I\hspace{-.1em}I}  &  3.5 & 50.08   &   333.133 &    -0.427 & 2008-09-04 & 5200121.1 &  IRCZ4 c;Nh & NG and NP \\
G338.398+0.164  & GH\,{I\hspace{-.1em}I}  & 13.1 & 50.90   &   338.358 &     0.154 & 2009-09-07 & 5200942.1 &  IRCZ4 c;Nh & NG and NP \\
G338.398+0.164  & GH\,{I\hspace{-.1em}I}  & 13.1 & 50.90   &   338.360 &     0.155 & 2009-09-07 & 5200942.2 &  IRCZ4 c;Nh & NG and NP \\
G338.400-0.201  & GH\,{I\hspace{-.1em}I}  & 15.7 & 50.24   &   338.409 &    -0.202 & 2009-09-07 & 5200943.2 &  IRCZ4 c;Nh & NG and NP \\
G345.528-0.051  & UNKNOWN  &  -- &  --   &   345.528 &    -0.049 & 2008-09-11 & 5200134.1 &  IRCZ4 c;Ns & NG and NP \\
G345.528-0.051  & UNKNOWN  &  -- &  --   &   345.530 &    -0.053 & 2008-09-11 & 5200133.1 &  IRCZ4 c;Nh & NG and NP 
\enddata 
\tablecomments{
{\dag$_1$}:\,Quoted from \cite{crow03,cont04}. 
{\dag$_2$}:\,See details in \citep{onak07}.
}
%\tablenotetext{$\dagger1$}{%
%}
\end{deluxetable} 
%ttttttttttttttttttttttttttttttttttttttttttttttttttttttttttttttttttttttttt%

%ttttttttttttttttttttttttttttttttttttttttttttttttttttttttttttttttttttttttt%
\begin{deluxetable}{lccc} 
\tablecolumns{4} 
\tablewidth{0pt} 
\tablecaption{Gaussian profile parameters for ice features\label{tab:2}}
\tablehead{
       \colhead{Band}  
    & \colhead{$\lambda_{k'_g}$ [$\mu$m]}
    & \colhead{$\gamma_{k'_g}$ [$\mu$m] for ``Ns''
 slit}
    & \colhead{$\gamma_{k'_g}$ [$\mu$m] for ``Nh''
 slit} 
    }  
\startdata 
CO$_2$ Ice & 4.26 & 0.060  & 0.048 \\
CO Ice         & 4.67 & 0.031  & 0.025  
\enddata 
%\tablecomments{
%}
%\tablenotetext{$\dagger1$}{%
%}
\end{deluxetable} 
%ttttttttttttttttttttttttttttttttttttttttttttttttttttttttttttttttttttttttt%

%ttttttttttttttttttttttttttttttttttttttttttttttttttttttttttttttttttttttttt%
\begin{deluxetable}{ccc} 
\tablecolumns{3} 
\tablewidth{0pt} 
\tablecaption{Lorentzian and Gaussian profile parameters for UIR bands\label{tab:3}}
\tablehead{
       \colhead{$\lambda_{k_l}$ [$\mu$m]}  
    & \colhead{$\gamma_{k_l}$ [$\mu$m] for ``Ns''
 slit}
    & \colhead{$\gamma_{k_l}$ [$\mu$m] for ``Nh''
 slit} 
    }  
\startdata 
 3.29 & 0.048  & 0.045 \\
 3.41 & 0.044  & 0.043  \smallskip \\ 
\hline\hline
$\lambda_{k_g}$ [$\mu$m] & $\gamma_{k_g}$ [$\mu$m]   for ``Ns''
 slit &  $\gamma_{k_g}$ [$\mu$m] for ``Nh''
 slit \\
\hline
 3.48 & 0.113 & 0.100  
 \enddata 
%\tablecomments{
%}
%\tablenotetext{$\dagger1$}{%
%}
\end{deluxetable} 
%ttttttttttttttttttttttttttttttttttttttttttttttttttttttttttttttttttttttttt%

%ttttttttttttttttttttttttttttttttttttttttttttttttttttttttttttttttttttttttt%
\begin{deluxetable}{lccc} 
\tablecolumns{4} 
\tablewidth{0pt} 
\tablecaption{Gaussian profile parameters for emission lines\label{tab:4}}
\tablehead{
       \colhead{Line}  
    & \colhead{$\lambda_{k_g}$ [$\mu$m]}
    & \colhead{$\gamma_{k_g}$ [$\mu$m] for ``Ns''
 slit}
    & \colhead{$\gamma_{k_g}$ [$\mu$m] for ``Nh''
 sllit} 
    }  
\startdata 
HI  Hu15                                                    & 3.91 & 0.031 & 0.025\\
HI  Hu13                                                    & 4.18 & 0.031 & 0.025\\
HI  Hu12                                                    & 4.38 & 0.031 & 0.025\\ 
HI  Pf12                                                     & 2.76 & 0.031 & 0.025\\
HI  Pf$\eta$                                               & 2.88 & 0.031 & 0.025\\
HI  Pf$\gamma$                                        & 3.75 & 0.031 & 0.025\\
HI  Pf$\beta$                                             & 4.65 & 0.031 & 0.025\\
HI   Br$\beta$                                            & 2.63 & 0.031 & 0.025\\
HI   Br$\alpha$                                          & 4.05 & 0.031 & 0.025\\
HeI  ($^3D_1-^3F_0$)                               & 4.30 & 0.031 & 0.025\\                      
H$_2$ 0|0 S(13)                                       & 3.85 & 0.031 & 0.025
\enddata 
\end{deluxetable} 
%ttttttttttttttttttttttttttttttttttttttttttttttttttttttttttttttttttttttttt%

%ttttttttttttttttttttttttttttttttttttttttttttttttttttttttttttttttttttttttt%
\begin{deluxetable}{ccc} 
\tablecolumns{3} 
\tablewidth{0pt} 
\tablecaption{Linear fit parameters for $I_{\rm 5.25\,\mu m}/I_{\rm 9\,\mu m}$ with $I_{\rm 3.3\,\mu m}/I_{\rm 9\,\mu m}$
%the 5.25\,$\mu$m band to the 9\,$\mu$m ratio with that of the 3.3\,$\mu$m band 
\tablenotemark{\,{\dag}}
\label{tab:5}}
\tablehead{
       \colhead{Objects}  
           & \colhead{$A_{\rm intercept}$} 
    & \colhead{$B_{\rm slope}$}
    }  
\startdata 
% \multicolumn{3}{l}{\bf the 3.3\,$\mu$m band versus the 5.25\,$\mu$m band}  \smallskip 
UCH\,{I\hspace{-.1em}I}+GH\,{I\hspace{-.1em}I}+UNKNOWN &  0.00007$\pm$0.00007 & 0.22$\pm$0.01\\
UCH\,{I\hspace{-.1em}I}                                                            &  0.00015$\pm$0.00012 & 0.24$\pm$0.01\\
GH\,{I\hspace{-.1em}I}                                                              &  0.00005$\pm$0.00008 & 0.22$\pm$0.01  
 \enddata 
\tablecomments{{\dag} see text. 
}
%\tablenotetext{$\dagger1$}{%
%}
\end{deluxetable} 
%ttttttttttttttttttttttttttttttttttttttttttttttttttttttttttttttttttttttttt%
%%%%%%%%%%%%%%%%%%%%%%%%%%%%%%%%%%
%%%%%%%%%%%%%%%%%%%%%%%%%%%%%%%%%%

\clearpage
%%%%%%%%%%%%%%%%%%%%%%%%%%%%%%%%%%
%%%%%%%%%%%%%%%%%%%%%%%%%%%%%%%%%%
%ttttttttttttttttttttttttttttttttttttttttttttttttttttttttttttttttttttttttt%
%ttttttttttttttttttttttttttttttttttttttttttttttttttttttttttttttttttttttttt%
  
%ttttttttttttttttttttttttttttttttttttttttttttttttttttttttttttttttttttttttt%
%ttttttttttttttttttttttttttttttttttttttttttttttttttttttttttttttttttttttttt%
  
%ttttttttttttttttttttttttttttttttttttttttttttttttttttttttttttttttttttttttt%
%ttttttttttttttttttttttttttttttttttttttttttttttttttttttttttttttttttttttttt% 

%ttttttttttttttttttttttttttttttttttttttttttttttttttttttttttttttttttttttttt%
%ttttttttttttttttttttttttttttttttttttttttttttttttttttttttttttttttttttttttt% 

%ttttttttttttttttttttttttttttttttttttttttttttttttttttttttttttttttttttttttt%
%ttttttttttttttttttttttttttttttttttttttttttttttttttttttttttttttttttttttttt% 

%ttttttttttttttttttttttttttttttttttttttttttttttttttttttttttttttttttttttttt%
%ttttttttttttttttttttttttttttttttttttttttttttttttttttttttttttttttttttttttt% 

%ttttttttttttttttttttttttttttttttttttttttttttttttttttttttttttttttttttttttt%
%ttttttttttttttttttttttttttttttttttttttttttttttttttttttttttttttttttttttttt% 

%ttttttttttttttttttttttttttttttttttttttttttttttttttttttttttttttttttttttttt%
%ttttttttttttttttttttttttttttttttttttttttttttttttttttttttttttttttttttttttt% 

%ttttttttttttttttttttttttttttttttttttttttttttttttttttttttttttttttttttttttt% 
%\include{Appendix_B}
%ttttttttttttttttttttttttttttttttttttttttttttttttttttttttttttttttttttttttt% 
%%%%%%%%%%%%%%%%%%%%%%%%%%%%%%%%%%
%%%%%%%%%%%%%%%%%%%%%%%%%%%%%%%%%%

\end{document}